\documentclass[aps,prb,twocolumn,showpacs,superscriptaddress,10p,longbibliography]{revtex4-1}

\usepackage{amsmath}
\usepackage{amssymb}
\usepackage[colorlinks=true,linkcolor=red]{hyperref}
\usepackage[sort&compress]{natbib}
\usepackage{graphicx}
\usepackage{color}

\begin{document}

\title{Frustration induced highly anisotropic magnetic patterns in classical $XY$ model on kagome lattice.}

\author{A.  Andreanov}
\affiliation{Center for Theoretical Physics of Complex Systems, Institute for Basic Science (IBS), Daejeon 34126, Republic of Korea}

\author{M. V. Fistul}
\affiliation{Center for Theoretical Physics of Complex Systems, Institute for Basic Science (IBS), Daejeon 34126, Republic of Korea}
\affiliation{Russian Quantum Center, National University of Science and Technology "MISIS", 119049 Moscow, Russia}

\date{\today}

\begin{abstract}
We predict and observed novel highly anisotropic  magnetic patterns obtained in  the model of frustrated planar interacting magnetic moments (the classical $X-Y$ model) on the regular kagome lattice. The frustration is provided by the presence of  both ferromagnetic and anti-ferromagnetic interactions between adjacent magnetic moments defined on the lattice nodes. At the critical value of the frustration $f=f_{cr}=3/4$ such a systems displays the phase transition from the ordered ferromagnetic state to the disordered frustration regime characterized by the highly-degenerated ground state. In the frustrated regime, $f_{cr}< f \leq 1$, unexpected scaling of spatially averaged magnetization $\langle \vec{M} \rangle $ on the total number of nodes,$N$, i.e. $\langle \vec{M} \rangle \simeq N^{-1/4}$, has been obtained. Such scaling is provided by highly anisotropic magnetic patterns displaying the ferromagnetic ordering along the $y$-direction, and short-range correlations of  magnetic moments along the $x$-direction. We conjecture that  all these intriguing features are explained  by the presence of the double-degenerated ground state in the basic cell, i.e. single triangle,  of the kagome lattice  accompanying a large amount of intrinsic constraints. We anticipate the implementation of the phase transition and  anisotropic magnetic patterns in various systems, e.g. natural magnetic molecular clusters, artificially prepared Josephson junctions networks, trapped-ions and/or  photonic crystals.
  
  %Frustration is induced by antifferomagnetic ... 
  %Such systems shows the phase transition from the homoggeneous %ferromagnetic phase to highly anisotropic ...
  %This papaern show short range correlation in a direction parallel %to the bonds and present a theoretical study of $\varphi_i$, i.e. %$C_p(i-j)=\langle\cos[p(\varphi_i - \varphi_j)]\rangle$ displaying %the transition from long- to short-range spatial correlations. We %find that higher-order correlations functions, e.g. $p=2$ and %$p=3$, restore the long-range behavior deeply in the frustrated %regime, $f\simeq 1$. Monte-Carlo simulations of the thermodynamics %of frustrated arrays of Josephson junctions are in good agreement %with analytical results. We also outline the extension of our %results to the case of kagome lattice, prototypical $2$D frustrated %lattice, and other higher dimensional lattices.
%\mikhail{The extension of our analysis to frustrated kagome lattice is also discussed. }
\end{abstract}
%\pacs{42.50.-p,74.81.Fa,74.50.+r}

\maketitle

Low-dimensional strongly interacting magnetic systems defined on the regular lattice fascinate scientists for many years. In such systems various magnetic phases, e.g. ferromagnetic and anti-ferromagnetic states, topological magnetic vortices and magnetic vortex-(anti)vortex pairs \cite{jos201340,jose1977renormalization},  and magnetic skyrmions \cite{nagaosa2013topological}, just to name a few, and corresponding phase transitions have been predicted and observed. Since the elaborated seminal mathematical models describing these phases are generic ones, analogous phases have been implemented and observed also in various artificially prepared solid states and optical systems, e.g. Josephson junction networks \cite{rzchowski1997phase,pop2008measurement,johnson2011quantum,king2018observation}, trapped-ions simulators \cite{britton2012engineered} and/or photonic crystals \cite{weimann2016transport,vicencio2015observation}. 

In this field a special interest presents a system of strongly interacting planar magnetic moments defined on various quasi-1D or 2D lattices, so-called \textit{classical $X-Y$ model}, showing both spin waves and topological excitations,  and the famous BKT transition \cite{jos201340,jose1977renormalization}. Even more peculiar inhomogeneous magnetic patterns have been obtained in the ground state of \textit{frustrated } magnetic lattices \cite{moessner2006geometrical,richter2018thermodynamic,baniodeh2018high}. E.g. the vortex states, the checkerboard distribution of vortices \cite{rzchowski1997phase,vu1993imaging}, strip phases \cite{valdez2005superconductivity} etc., and sharp transitions between these magnetic patterns as the external parameter varies,   were observed. The frustration in regular magnetic systems can be provided by various means: application of an external magnetic field; fabrication of specific lattices such as honeycomb, Lieb or kagome lattices; combinations of periodically distributed magnetic couplings having alternating signs (ferromagnetic/anti-ferromagnetic interactions).
%We are using a different definition because frustration arises due %to the
%Josephson couplings
%having alternating signs in a single lattice cell in our model. %between adjacent magnetic moments. 
Prediction of magnetic patterns in frustrated magnetic systems is rather sophisticated problem, there are just a few models where complete analytical solutions are known. 

In this Letter we present an analytic solution of the particular frustrated classical $X-Y$ model on the 2D kagome lattice. Our analysis is heavily based on a peculiar geometrical property of the kagome lattice, namely, the vertex-shearing  triangles \cite{moessner2006geometrical}, i.e.  the basic cells (triangles) of the lattice are connected through the nodes (see Fig. 1) but not the edges as it is e.g. for square lattices. Here, we will use a rather new way of introducing frustration, in which frustration  arises due to the periodically distributed magnetic couplings having alternating signs in a single lattice cell. A  great advantage of such method of introducing frustration is the persistence of the frustrated regime for a substantial range of frustrations but not a single one as it was discussed in overwhelming number of scientific works previously. 

These ideas have been applied in Ref. \cite{andreanov2019resonant} for the analysis of various inhomogeneous patterns in Josephson junction networks forming quasi-1D diamond and saw-tooth chains.  The two phases showing the ferromagnetic ordering and disordered pattern of penetrating vortices (antivortices), and corresponding phase transition between them have been found.  Here, we notice that the frustrated classical $X-Y$ model can be naturally implemented in artificially prepared Josephson junctions networks. 
However, experimental realizations of such arrays requires Josephson couplings of different signs. Such Josephson couplings are provided by the so-called $\pi$-Josephson junctions that can be
fabricated on basis of multi-junctions SQUIDS
in externally applied magnetic field  \cite{rzchowski1997phase,pop2008measurement,johnson2011quantum,king2018observation,jung2014multistability,shulga2018magnetically}, superconductor-ferromagnet-superconductor junctions \cite{feofanov2010implementation}, different
facets of grain boundaries of high temperature superconductors \cite{hilgenkamp2008pi}, Josephson junctions between two-bands superconductors \cite{dias2015origami}.

As we turn to the frustration regime of $X-Y$ model on 2D   kagome lattice  novel highly-anisotropic magnetic patterns are predicted and obtained. These magnetic patterns will be characterized by spatial correlation functions. The fingerprints of these magnetic patterns is the dependence of the spatially averaged magnetization on the system size showing unusual scaling. The highly-degenerated ground state occurring in the frustration regime, will be studied in detail, and all possible realizations of such ground state will be classified.    

Let us consider the model of interacting classical planar magnetic moments defined on the 2D kagome lattice of a finite size $N$, where $N$ is the total number of nodes in the lattice. In this model each node $i$ of the lattice is characterized by the two-component magnetic moment, 
$\vec{M}_i=\{M_{x,i}, M_{y,i} \}$, where $(x,y)$ are axis in the plane of the lattice. The thermodynamic properties of such a system are determined by the interaction part of the Hamiltonian written as :
\begin{equation}
    \label{SpinHamiltonian}
   H=-J\sum_{ij}\alpha_{ij}\vec{M}_i \vec{M_j},
\end{equation} 
 where $J\alpha_{ij}$ is the interaction strength. We assume that the interaction occurs between the magnetic moments of adjacent nodes $i$ and $j$ only, and therefore, $\alpha_{ij}$ yield not zero values on the edges of the kagome lattice. Next, we introduce the frustration in our model as follows: $\alpha_{ij}=1$ for all edges except the "horizontal", i.e. along the $x$-direction, ones; $\alpha_{ij}=\alpha$ for horizontal edges. Moreover, the interaction strength $\alpha$ can take negative values, and, in general case, $\alpha$ varies from $1$  to $-1$. We define also the frustration parameter $f=(1-\alpha)/2$, where the region of frustrations,  $0\leq f \leq 0.5$,   corresponds to the presence of  ferromagnetic interactions  only,  and in the range of frustrations, $0.5<f \leq 1$, there are both ferromagnetic and anti-ferromagnetic couplings periodically varying in space. The schematic of interacting planar magnetic moments on the kagome lattice showing various magnetic moments patterns for different values of frustration $f$ are presented in Fig. 1. 
 
 Another way to describe such a system is to define the phases $\varphi_i$ on the nodes as $M_{x,i}=M\cos \varphi_i$ and $M_{y,i}=M\sin \varphi_i$  ($M$ is the amplitude of the magnetic moments), and in this representation, the Hamiltonian is written as \cite{jose1977renormalization,korshunov2004fluctuations,korshunov2005fluctuation,rzchowski1997phase}

\begin{equation}
    \label{PhaseHamiltonian}
   H\{\varphi_i\}=-MJ\sum_{ij}\alpha_{ij}\cos(\varphi_i -\varphi_j).
\end{equation} 
Such phase-Hamiltonian (Eq. (\ref{PhaseHamiltonian})) is extremely suitable for an analysis of the frustrated planar magnetic moment models, and  it also describes the dynamics of complex Josephson junction networks composed of both $0$- and  $\pi$ junctions \cite{feofanov2010implementation,hilgenkamp2008pi,dias2015origami}. For Josephson junction networks the parameter $MJ$ has a meaning of the critical current of a single Josephson junction between the superconducting islands occupied the lattice nodes.

\begin{figure}
    \includegraphics[width=1\columnwidth]{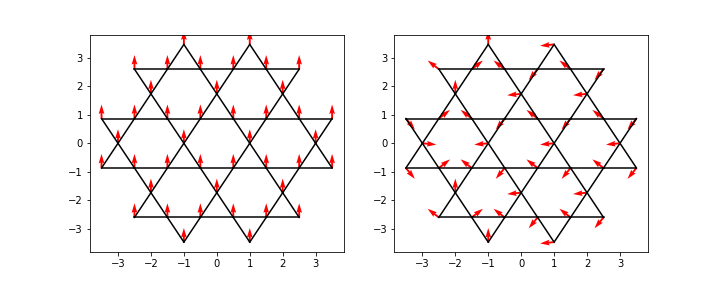}
    \caption{(color online) The schematics of  planar magnetic moment model (the classical $X-Y$ model on the kagome lattice) in non-frustrated (a) and frustrated (b) regimes. The ferromagnetic  ((a), $f=0$) and disordered magnetic moment pattern ((b),$f=1$) are shown} 
   \label{States}
\end{figure}

At temperature $T=0$ the spatially averaged  magnetization $\langle M \rangle $, and various magnetic patterns are determined by the  ground state of a system. By making use of a particular property of the kagome lattice, namely, the vertex-shearing triangles, 
%basic cells (the triangles shown in Fig.1) are connected through %the nodes but not edges, 
we obtain the  ground states of a single triangle minimizing of its energy. With obtained ground states of a single triangle we were able to obtain the ground states of a whole system for different values of the frustration $f$. We also numerically calculate spatially averaged magnetization $\langle M \rangle=\frac{1}{N}|\sum_i \vec{M}_i|$ , and spatial correlation functions, $C(i-j)=\langle \vec{M}_i \vec{M}_j \rangle$. 

The dependence of spatially averaged magnetization $\langle M \rangle$ on the frustration $f$ is shown in Fig. 2a. We obtain that in the region of frustrations, $0<f<f_{cr}=3/4$ , the ground state is the ferromagnetic one. In this non-frustrated regime the spatially averaged magnetization does not depend on $N$ and $f$, and takes a  maximum value, $\langle M \rangle=M $.  The stability of the ferromagnetic ground state was obtained by studying small oscillations around the ground state \cite{richard1972feynman}. To make that we write the dynamic equations of the phase-Hamiltonian (\ref{PhaseHamiltonian}) in the following form: 
\begin{equation}
    \label{PhaseDynamics}
   \ddot \varphi_i=-\frac{\partial H}{\partial \varphi_i}.
\end{equation} 
Small collective oscillations of phases $\varphi_i$ are characterized by the dispersion relation $\omega(k)$. In the Ref. \onlinecite{andreanov2019resonant} for $f=3/4$ the zero mode $\omega(k)=0$ has been obtained. Such flat band zero mode indicates the instability of the ferromagnetic ground state $\vec{M}_i=\vec{M}$ ($\varphi_i=0$), and the phase transition to the \textit{intrinsically disordered frustrated regime} occurring in the region of frustrations, $(3/4)<f<1$.

In this regime the spatially averaged magnetization $\langle M \rangle$ strongly depends on the frustration $f$ and size of a system, $N$. In the limit of $N \rightarrow \infty$ the $\langle M \rangle$ aspires the zero value, and it indicates the presence of disordered magnetic patterns. The quantitative analysis of disordered magnetic patterns was carried out as follows:
$$ 
\langle M \rangle=\sqrt{ \overline{M^2}}
$$
\begin{equation}
    \label{Magnetization-1}
   \overline{M^2}=\frac{1}{N^2}\left [\left (\sum_i M_{x,i} \right )^2+\left (\sum_i M_{y,i} \right)^2 \right ] . 
\end{equation}
For systems of a large size $N \gg 1$ we  use the continuous limit, and introducing the spatial correlation function $C(\vec{\rho})$, where $\vec{\rho}$ is the radius vector in the lattice plane, the spatially averaged magnetization is expressed as 
\begin{equation}
    \label{Magnetization-2}
   \overline{M^2}=\frac{M^2}{L^D}\int d^D\vec{\rho}C(\vec{\rho}),  
\end{equation}
where $L \simeq N^{1/D}$ is the linear size of a system, and $D$ is the dimension of the lattice. Thus, the dependence of $\langle M \rangle$ on the size $L$ (or number of sites $N$) is completely determined by the spatial correlation function $C(\vec{\rho})$.

Assuming an isotropic correlation function with the finite correlation radius $\rho_0(f)$, we obtain $\langle M \rangle \propto 1/\sqrt{N}$ as $L \gg \rho_0$.  First, we apply this result to the case of 1D frustrated saw-tooth chain where the short range correlation function has been obtained \cite{andreanov2019resonant}.  Indeed, the scaling $\langle M \rangle \propto M \rho_0/\sqrt{N}$ was observed for frustrated saw-tooth chain (see the dashed line in Fig. 2b). However, numerically calculated dependencies of $\langle M \rangle$ on the size of the kagome lattice clearly demonstrate unexpected  scaling 
$\langle M \rangle \propto N^{-1/4} $ (see the solid lines in Fig. 2b).
Such scaling indicates a strong anisotropy of the correlation function
$C(\vec{\rho})$ for the magnetic kagome lattice biased in the frustrated regime. Indeed, the scaling $N^{1/4}$ can be explained by a following assumption: in the frustrated regime the spatial correlation function shows the long-range correlation in a one direction, and  a short range correlation in other direction. 

\begin{figure}
    \includegraphics[width=1\columnwidth]{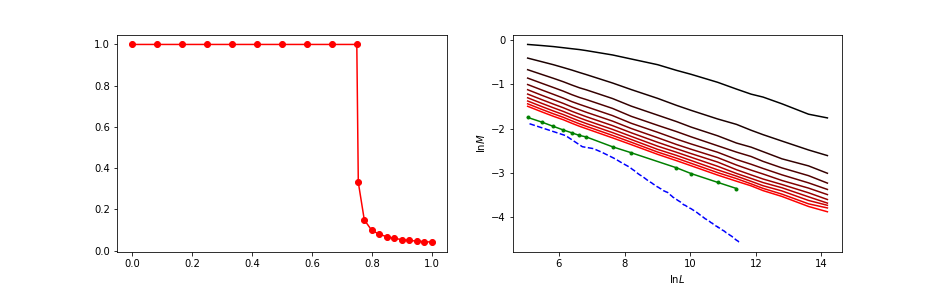}
    \caption{(color online) The dependence of spatially averaged magnetization $\langle M \rangle $ on the frustration $f$ (a) and the size of the system $N$ (b) (solid lines). The size $N=10^4$ was chosen for Fig. 2a. In Fig. 2b the frustration $f$ decreases from bottom ($f=1$) to up.  The dependence of spatially averaged magnetization $\langle M \rangle $ on  the size of the system $N$ for 1D frustrated saw-tooth chain is shown in (b) by dashed line. } 
   \label{Magnetization}
\end{figure}

To check this assumption we numerically calculated the spatial correlation function $C(\vec{\rho})$ for different values of frustration. The typical dependencies are presented in Fig. 3 for $f=4/5$ and for two different sizes of a system, and one can see an anisotropic magnetic pattern with short range  correlations along the $x$-direction (see the $C_x(\rho)$ in Fig. 3a), and a long-range ferromagnetic correlation along the $y$-direction (see the $C_y(\rho)$ in Fig. 3b). Notice here that for systems of moderate size ($N \simeq 250$), the long-range ferromagnetic correlations diminish (see red (grey) line in Fig. 3b).  The correlation radius of 
$C_x(\rho)$ increases as the frustration $f$  approaches to the critical value $f_{cr}$.
\begin{figure}
    \includegraphics[width=1\columnwidth]{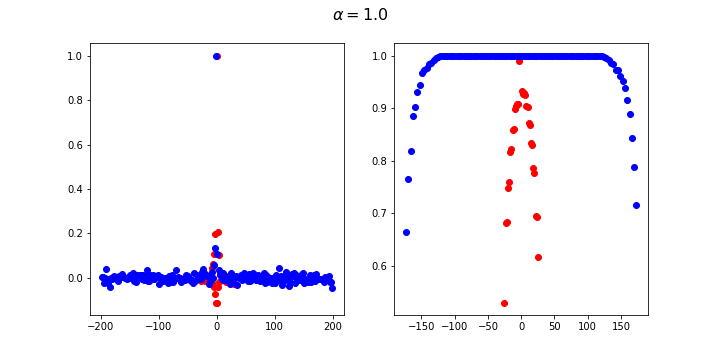}
    \caption{(color online) The anisotropic spatial correlation functions  of the magnetic moments $C_{x,y}(i-j)$ in the frustrated regime: long-range ferromagnetic correlations along the $y$-direction (a), short-range correlations along the $x$-direction (b). The results for the frustration $f=4/5$, and two different sizes $N=250$ (red (gray) lines) and $N=6 \times 10^4$ (blue (light gray) lines) are presented. } 
   \label{CorrelationFunction}
\end{figure}

Next, we present qualitative explanation of such anisotropic magnetic patterns occurring in the frustrated (disordered) regime of the $X-Y$ model on the kagome lattice. The thermodynamic properties of the classical $X-Y$ model on the kagome lattice is determined by the partition function

In the classical regime the thermodynamic properties of the arrays of interacting Josephson
junctions are described by the partition function
\begin{equation}
    \label{PartitionFunction}
 Z=\int D\{\varphi_i\} exp \left [\frac{H \{\varphi_i \}}{k_B T}\right]
\end{equation}
At $T=0$ the partition function $Z$ is determined by the global minimum of the functional $H \{\varphi_i\}$ only. As $f<f_{cr}$ such minimum corresponds to the ferromagnetic ordering, $\varphi_i=0$. As $f>f_{cr}$ the minimum  of Hamiltonians (\ref{SpinHamiltonian}) and (\ref{PhaseHamiltonian}) of \textit{a single triangle}, is provided by two realizations of the phases $\varphi_{1-3}$ (these phases are indicated in Fig. 4a). Therefore, the ground state of a single triangle is  the double-generated one, and explicit values of the phases $\varphi_{1-3}$ were obtained in Ref.  \onlinecite{andreanov2019resonant} as $\varphi_2-\varphi_1=\varphi_3-\varphi_2=u$, and $\varphi_3-\varphi_1=2u$. The values of $u$ are determined by the  frustration as $u=\pm 2 \arccos [\frac{1}{(4f-2)}]$ in the range of frustrations, $f_{cr}<f \leq 1$, and two realizations correspond to the penetration of vortex (anti-vortex)  that differ by clockwise (anticlockwise)  rotation of the magnetic moments $\vec{M}_{1-3}$. The vortices (antivortices) are  marked by (green)red circles in Fig. 4a-b. 

In the frustration regime $f_{cr}<f \leq 1$ we construct the ground states of a kagome lattice taking into account all realizations of these two basic states of a single triangle, satisfying  a large amount of intrinsic constraints: the sums of phases $\varphi_i$ around the closed loops have to be $2\pi n$. For most frustrations just the value $n=0$ can be realized. 

Thus, a single plaquette of the kagome lattice composed of six triangles and a one closed loop, provides the 14th-degenerated ground state corresponding possible combinations of (anti)vortices The seven realizations of the ground state are shown in Fig. 4c-j, and other seven realizations can be obtained by exchanging red (green) marks. Notice here that the total number of randomly distributed red (green) circles is $2^6=64$, and therefore, even a single constraint  drastically reduces the number of possible realizations. 

As we turn to kagome lattices of a large size  (a particular example with 24 triangles and seven constraints is shown in Fig.1) we also obtain that a number of realizations $R$ corresponding to the highly-degenerated ground state is  drastically reduced, and our study indicates the following  conjecture: 
$log_2 R \simeq \sqrt{N}$ instead of $log_2 R \simeq N$ for randomly distributed (anti)vortices. Moreover, the overwhelming number of realizations shows the ferromagnetic ordering along the $y$-direction, and no such ordering in $x$-direction. It results in the anisotropic spatial correlation functions $C_{x,y}(i-j)$ averaged over the realizations corresponding to the highly degenerated ground state (see Fig. 3). However, such ferromagnetic ordering along the  $y$-direction was not observed for systems of moderate size (see red ( gray) solid lines in Fig. 3). 

Finally, we notice that the additional realizations corresponding to the  sums of phases $\varphi_i$ around the closed loops have to be $\pm 2\pi$, were obtained for particular values of frustrations as $f=1$ and $f=(\sqrt{3}+3)/6$. For example for the case $f=1$ these additional realizations lead to the recovering of isotropic spatial correlation function $C_{x,y}(i-j)$. However, as $f=1$ the spatial isotropic correlation function shows long-range correlations as $C(\rho) \simeq 1/\rho$, and the spatially averaged magnetization still displays the scaling $\langle \vec{M} \rangle \simeq N^{-1/4}$ (see Fig. 2b).

\begin{figure}
    \includegraphics[width=1\columnwidth]{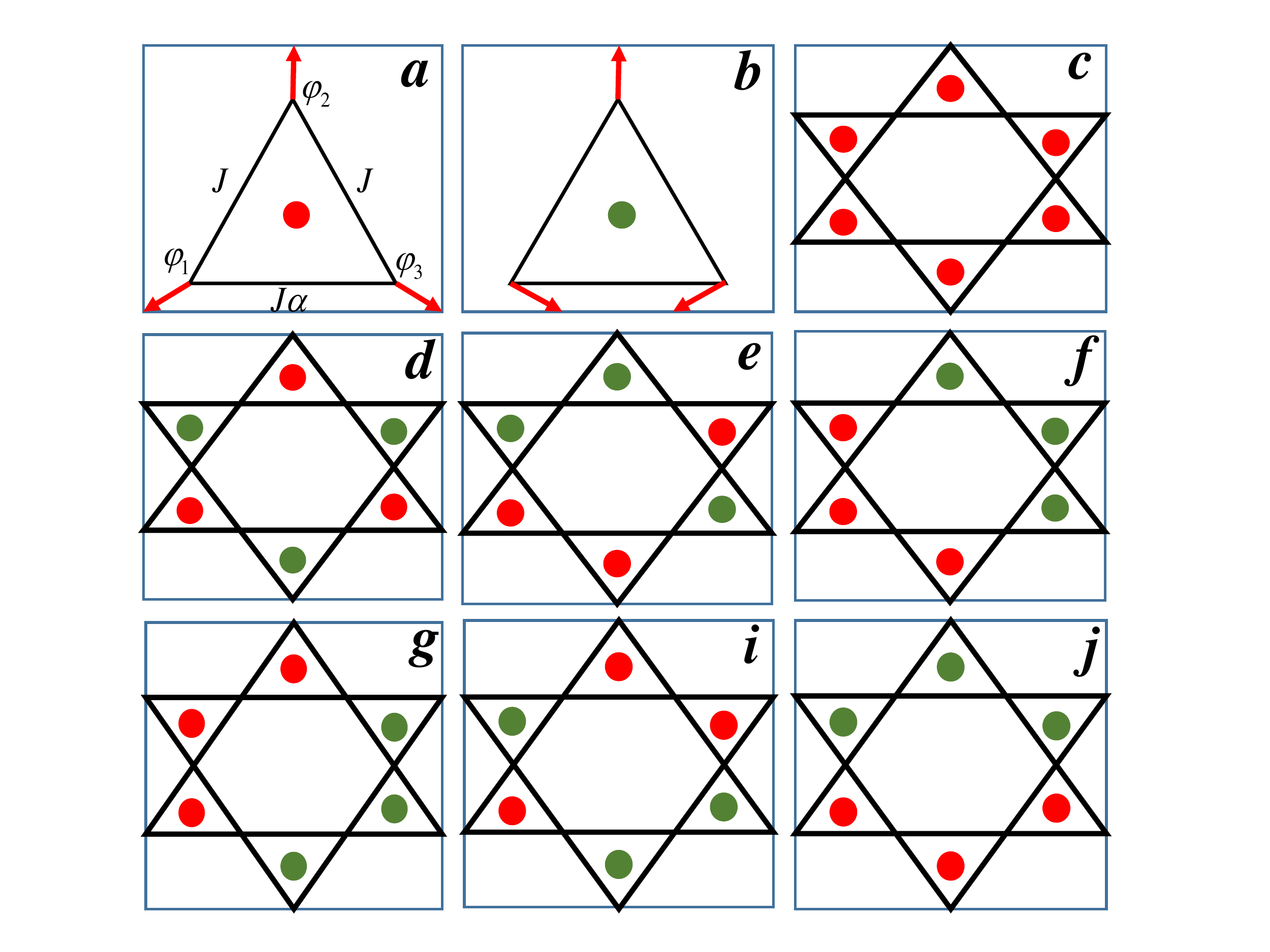}
    \caption{(color online) Distributions of magnetic moments in a single triangle (a-b), and a single plaquete of the kagome lattice (c-j). The (anti)clockwise distributions of a single triangle magnetic moment $\vec{M}_{1-3}$ are marked by red(green) circles. The phases $\varphi_{1-3}$ on nodes and interaction strengths are indicated in Fig. 4a.}
   \label{Magnetization}
\end{figure}

In conclusion, we have studied various magnetic patterns in the model of frustrated planar interacting magnetic moments (the classical $X-Y$ model) on the kagome lattice. The frustration is provided by the presence of both ferromagnetic and anti-ferromagnetic interactions between adjacent magnetic moments. At the critical value of the frustration $f=f_{cr}=3/4$ such a systems shows the phase transition from the ordered ferromagnetic state to the disordered frustration regime characterized by highly-degenerated ground state (see Fig. 2a). In this frustrated regime, $f_{cr}< f \leq 1$, unexpected scaling of spatially averaged magnetization $\langle \vec{M} \rangle $ on the number of nodes, $N$, i.e. $\langle \vec{M} \rangle \simeq N^{-1/4}$, has been obtained (see Fig. 2b). Such scaling is determined by the anisotropic magnetic patterns displaying the ferromagnetic ordering along the $y$-direction, and short-range correlations of  magnetic moments along the $x$-direction (see Fig. 3). All these intriguing features were explained by the presence of double-degenerated ground state of a basic cell, i.e. single triangles, of the kagome lattice (see Fig. 4a-b) accompanying a large amount of intrinsic constraints (see Fig. 4c-j). The classical frustrated $X-Y$ model on the kagome lattice can be implemented in natural magnetic molecular clusters, artificially prepared Josephson junctions networks, and  photonic crystals, and we anticipate the observation of the phase transition and various anisotropic magnetic patterns in such systems.  

\textbf{Acknowledgments}
The authors thank S. Flach for useful discussions. This work was supported by the Institute for Basic Science in Korea (IBS-R024-D1). M. V. F. acknowledges the partial financial support of the Russian Science Foundation in the framework of the  Project $19-42-04137$. 

\bibliography{general,flatband,josephson}

\end{document}